\documentstyle[prl,aps,multicol,epsfig]{revtex}

\begin{document}
\draft \preprint{}
\title{Theory of Coherent and Incoherent Nuclear Spin-Dephasing in the
Heart}

\author{Wolfgang R. Bauer*\cite{bylineWolfgang},
Walter Nadler**, Michael Bock ***, Lothar R. Schad ***, \\
Christian Wacker*, Andreas Hartlep**, Georg Ertl*}

\address{
* II. Medizinische Klinik,
Klinikum Mannheim / Universit\"at Heidelberg, \\ Theodor-Kutzer
Ufer 1-3, D-68167 Mannheim, Germany, }

\address{
** HLRZ/NIC c/o Forschungszentrum J\"ulich, D-52725 J\"ulich,
Germany }

\address{
*** Department of Biophysics, German Cancer Research Center
(DKFZ), D-69120 Heidelberg, Germany }

\date{\today}

\maketitle

\begin{abstract}

We present an analytical theory of susceptibility induced nuclear
spin dephasing in the capillary network of myocardium. Using a
strong collision approach, equations are obtained for the
relaxation rate of the free induction and the spin echo decay.
Simulation and experimental data are well predicted by the theory.
Since paramagnetic deoxyhemoglobin as the origin of nuclear spin
dephasing has a higher tissue concentration in myocardium supplied
by a stenotic, i. e. significantly narrowed, coronary artery, spin
dephasing might serve as a diagnostic tool. Our approach can be
modified for capillary networks in other tissues than myocardium
and may be applied in material science.

\end{abstract}

\pacs{ 87.59.Pw, 76.60.Jx }

\begin{multicols}{2}

\narrowtext

In cardiology a paramount goal is the detection of significant
narrowing, called stenosis, of coronary arteries, and -- even more
important -- the detection of the associated area of the cardiac
muscle. There is evidence that this area can be distinguished from
normally supplied tissue without contrast agent by NMR imaging
from its different transverse spin relaxation time \cite{wacker}.
Water proton spin dephasing is induced by spin-spin interactions
and susceptibility differences between different  histological
substructures such as intra- and extravascular space. In native
tissue this difference is related to the paramagnetic property of
deoxygenated hemoglobin which induces perivascular field
gradients. Deoxyhemoglobin which results from intracapillary
dexoygenation of almost 100\% oxygenated arterial hemoglobin, is
predominately located in the capillary and venous system. In the
cardiac muscle this blood oxygen level dependent (BOLD) effect is
almost solely due to the magnetic field inhomogenties around
capillaries since they contain more than 90\% of intramyocardial
blood volume \cite{kaul}. In myocardium supplied by a stenotic
coronary artery, coronary autoregulation tries to maintain
sufficient perfusion by compensatory filling of capillaries
(recruitment), i.e. deoxyhemoglobin increases in tissue and spin
dephasing is accelerated when compared to normally supplied tissue
\cite{wacker}. This difference of spin dephasing can be pronounced
by pharmacological coronary dilation since a higher oxygen supply
decreases the tissue concentration of deoxyhemoglobin in normal
myocardium \cite{li,rosen,wacker2} whereas it does not in the
poststenotic myocardium observed \cite{wacker}.

Though the relation between spin dephasing and coronary
microcirculation is of great interest for the diagnosis of
coronary artery disease, at present there is no satisfying
analytical theory that describes the dephasing of diffusing spins
in the inhomoegneous magnetic field of the capillary system. There
exist theories on spin dephasing for very long and very short
correlation times of diffusion induced magnetic field
fluctuations. The assumption of static dephasing of spins as
proposed by Yablonski et al. \cite{yablonski} is not valid in
human cardiac NMR imaging systems (field strength less than 2
Tesla) since the diffusion distance $l$ during relaxation
(relaxation time $T_2^*>30$ms, $D=1{\mu{\rm m}^2/{\rm ms}} $,
$l>\sqrt{2 T_2^* D}\approx 7\mu$m) has the same magnitude as the
intercapillary distance ($19\mu$m). Thus, a single nuclear spin
experiences almost the whole pericapillary magnetic field gradient
during its relaxation. Kisselev and Posse \cite{posse} recently
proposed analytical treatments for long and short correlation
times. The first extends the static dephasing model by including
diffusion. However, only the effect of a {\it linear} local field
gradient on relaxation is considered. Application of this approach
would be valid if large vessels would dominate the microvascular
architecture of myocardium, which is, however, not the case. The
other is a perturbation approach in the local fields. In the
cardiac muscle the motional narrowing condition
$\tau\langle\omega^2\rangle^{1/2}\ll 1$ with correlation time
$\tau$ and variance of fluctuations of the local fields (expressed
as frequency) $\langle\omega^2\rangle$ is not fulfilled, though.
Approaches which consider spin dephasing for arbitrary correlation
times as the Gaussian or Lorentzian phase approximation were not
sucessful since they neither revealed the correct field dependence
of the relaxation rate \cite{kennan}, nor the correct
fluctuation-dissipation relation \cite{posse}.

In this paper an analytical theory is proposed which describes
coherent and incoherent spin dephasing in the capillary network of
myocardium. It is not an extension of the static dephasing model
or a perturbation approach. Instead, it is based on a strong
collision approximation which includes both, the static dephasing
and the motional narrowing regime as two limiting cases.

It has to be emphasized that this approach is not restricted to
myocardium, but can be modified for the description of transverse
relaxation in capillary networks of other tissues as well.
Applications in material science are possible, too. For example,
the effect of complex potentials on spin diffusion, and, hence, on
spin dephasing, is exploited to determine the microstructure of
spatially or magnetically heterogeneous media \cite{callaghan}.
Examples for the former are porous structures, for the latter
polymer solutions and liquid crystalls in which magnetic
heterogenity is induced by either paramagnetic inpurities or an
inhomogeneous electrical current. After modifications for the
appropriate geometry the analysis presented below should also be
applicable in that field of research.

We will first discuss the {\it tissue model} employed. The regular
parallel architecture of muscle fibers and capillaries in
myocardium \cite{bassingthwaighte} allows to reduce considerations
from the whole tissue to one tubular capillary (radius $R_c$)
which is concentrically surrounded by its mean cylindrical supply
region with radius $R_s$ (Krogh model  \cite{krogh}). Nuclear spin
diffusion in the whole tissue is replaced by restricted diffusion
in the supply region, i.e. reflectory boundary conditions are
introduced at $R_s$, and, since the capillary wall is  nearly
impermeable for water on the time scale of spin dephasing
\cite{donahue}, also at $R_c$. We only consider relaxation of
extracapillary nuclear spins since in myocardium the relative
intracapillary blood fraction $\varsigma=R_c^2/R_s^2$ is less than
10\%. The effect of diffusion in field gradients along the
capillary axis on spin dephasing is negligible, i.e. it is only
necessary to consider two-dimensional diffusion of spins within
two concentric circles ($R_c$, $R_s$). Basic magnetostatics
provides the susceptibility related spin frequency shift around
the capillary in cylindrical coordinates as
\begin{equation}
\label{dipole} \omega(r,\phi)=-\delta\omega R_c^2\cos(2\phi)/r^2\;
,
\end{equation}
with the characteristic equatorial frequency shift
$\delta\omega=2\pi\gamma\Delta \chi B_0\cos(\theta)$ where
$\gamma$ is the gyromagnetic ratio, $\Delta\chi$ the susceptibilty
difference and $\theta$ is the tilt angle between the capillary
axis and the external field which is almost $90^o$ in a clinical
scanner. Since$\Delta\chi<8\cdot10^{-8}$ for blood the
intracapillary magnetization is less than 1 mgauss for clinical
NMR magnets and $\delta\omega\le 168$ rads$^{-1}$. The
contribution of fields of neighbouring capillaries is negligible
due to the intercapillary distance of 19$\mu$m.

The starting point of our analytical treatment is the free
induction decay of the local nuclear transverse magnetization. The
time evolution (polar notation: $m=m_x-im_y$) at $x=(r,\phi)$  is
determined according to the Bloch Torrey diffusion equations
\cite{torrey} which, written in rotating frame coordinates, are
\begin{equation}
\label{torreyg}
\partial_t m(x,t)=(D \nabla^2 + i\omega(x))\; m(x,t)\; ,
\end{equation}
with diffusion coefficient $D$. The formal solution of
Eq.~(\ref{torreyg}) is $m(x,t)=\exp[(D \nabla^2 + i\omega(x))t]
m(x,0)$. Since the spatial resolution of clinical NMR scanners is
far above that of capillary dimensions (mm $vs$ $\mu$m) the
observable is the mean transverse magnetization $M(t)$ within the
supply region with volume $V$
\begin{equation}
\label{torreyi} M(t)=V^{-1} \int_V dx\; \exp[(D \nabla^2 +
i\omega(x)) t]\; m(x,0)\; .
\end{equation}
The homogeneous water concentration in the cardiac muscle implies
a homogeneous initial magnetization distribution $m(x,0)$ which
for simplicity is normalized to $m(x,0)=1$. Since
Eq.~(\ref{torreyg})cannot be solved analytically when the local
frequency is given by Eq.~(\ref{dipole}), a different approach is
necessary.

Let us first consider the {\it fluctuations} of the local magnetic
fields, represented by the corresponding local NMR-frequencies
$\omega(x)$, that a diffusing nuclear spin experiences. Their
autocorrelation function $K(t)$ is determined by
$
K(t)=V^{-1} \int_V dx\; \omega(x)\exp(t D\nabla^2)\omega(x) $,
where $\exp(t D\nabla^2)$ is the Green's function of the
restricted diffusion. The {\it correlation time} $\tau$ of the
field fluctuations is defined as the mean relaxation time
\cite{nadler} of the autocorrelation function,
 $\tau=\int_0^{\infty} dt K(t)/K(0)$.
Using the techniques of Ref.~\cite{nadler}, we obtain
\cite{bauer99a}
\begin{equation}
\label{tau} \tau=\left(R_c^2/4D\right)\ln(\varsigma
)/(\varsigma-1)\; .
\end{equation}

Inserting realistic values of relative intracapillar blood volume
($\varsigma=5$ to $10\%$), capillary radius ($R_c=2.5$ to
$2.75\mu$m) and the diffusion coefficient ($D=1{\mu{\rm m}^2/{\rm
ms}}$) into Eq.~(\ref{tau}) implies that $\tau\le 6 {\rm ms}$.
Relaxation times of the free induction decay observed using
clinical scanners are $T_2^*\ge 30 {\rm ms}$. Hence, the
relaxation time describing the contribution of field
inhomogenities is even longer. This implies that on a time scale
in which significant variations of $M(t)$ occur the local field
fluctuations are stochastically independent.

This stochastic independence suggests replacing the diffusion
operator $D\nabla^2$ in Eq.~(\ref{torreyg}) by a {\it strong
collision operator} \cite{dattagupta}. This operator describes the
diffusion as a stationary Markov process with a single rate
$\lambda$ that governs the relaxation to the steady state
distribution $p(x)$, independent of the initial state. Due to the
homogeneity of water in tissue the steady state distribution is
simply $p(x)=1/V$. The generator ${\bf D}$ of the corresponding
Markov process is then
\begin{equation}
\label{str} {\bf D}=\lambda(\bf{\Pi-id})\;
\end{equation}
where ${\bf\Pi}$ denotes the projection operator onto the
functional space generated by $p(x)$, i.e. the application of this
operator on some function $f(x)$ is,
 ${\bf\Pi}f(x)=V^{-1}\int_V dx\; f(x)\;$.
The constant $\lambda$ is determined selfconsistently by requiring
that ${\bf D}$ describes correctly the field fluctuations. This
implies $\lambda=\tau^{-1}$, see Eq.~(\ref{tau}).

Instead of directly solving the time evolution of the
magnetization, Eq.~(\ref{torreyi}), under the strong collision
assumption, it is more convenient to consider the Laplace
transform
\begin{equation}
\label{lpt1} \hat M(s) = V^{-1} \int_V dx\; \left[s - D \nabla^2 -
i\omega(x)\right]^{-1}\; m(x,0)\; .
\end{equation}
Replacing $D\nabla^2$  by ${\bf D}$ and utilizing the operator
identity $(A+B)^{-1}=A^{-1}-A^{-1}B(A+B)^{-1}$ we obtain after
some algebra
\begin{eqnarray}
\label{lpt2} \hat M(s) &=& (1+\varsigma) \left[
\sqrt{(s+\tau^{-1})^2+\delta\omega^2 \varsigma^2} + \right. \cr &
& \quad \quad \left.
\varsigma\sqrt{(s+\tau^{-1})^2+\delta\omega^2}-\tau^{-1}
(1+\varsigma)\right]^{-1} \; .
\end{eqnarray}
$M(t)$ can be obtained from this result to arbritrary accuracy as
a multi-exponential function $M(t)\approx\sum_{\nu=1}^N
f_\nu\exp(-\Gamma_\nu t)$, the parameters determined from
(\ref{lpt2}) using the generalized moment method of \cite{nadler}.
The relaxation time of the free induction process corresponds to
the lowest order (mean relaxation time) approximation, which
provides the best single exponential approximation, $M(t)\approx
e^{-t/T_2^*}$, by setting $T_2^*=\hat M(0)$. Thus,
\begin{equation}
\label{T2star} T_2^* = \frac{\tau (1+\varsigma)}{
[\sqrt{1+(\varsigma\tau\delta\omega)^2}-1] +
\varsigma[\sqrt{1+(\tau\delta\omega)^2}-1]}\; .
\end{equation}

The free induction decay of magnetization is a result of coherent
and incoherent spin dephasing. When instead a {\it spin echo}
experiment is performed, a $180^0$ pulse applied at $t/2$ reflects
the spin phase which eliminates the coherent contribution of spin
dephasing at the echo time $t$. Time evolution of the local
magnetization before the $180^0$ pulse is that of a free induction
decay $m(x,t/2)=\exp[(D \nabla^2+i\omega(x))t/2]m(x,0)$. After the
phase reflection which is described by the complex adjoint,
$m(x,t/2)\stackrel{180^0\;{\rm pulse}}{\longrightarrow}
m^*(x,t/2)$ there is again a free induction decay, i.e. the global
magnetization ${M_{SE}} (t)$ after the echo time $t$ is
\begin{eqnarray}
{M_{SE}}(t)&=& V^{-1}\int_V dV \; \exp[(D
\nabla^2+i\omega(x))t/2]\times \cr & & \quad\quad\quad \exp[(D
\nabla^2-i\omega(x))t/2 ] m(x,0) \quad .
\end{eqnarray}
Again replacing $D\nabla^2$  by ${\bf D}$, differentiating this
relation and exploiting the structure of the operator ${\bf D}$ we
obtain a general relation between the contribution of solely
incoherent spin dephasing on relaxation (${M_{SE}}$) and that of
coherent and incoherent dephasing ($M$)
\begin{equation}
\partial_t{M_{SE}}(t)
=\frac{1}{\tau} \left( \mid M(t/2)\mid^2-{M_{SE}}(t) \right) \;,
\label{SEGE}
\end{equation}
which is solved by
\begin{equation}
\label{sol} {M_{SE}}(t)=\frac{e^{-t/\tau}}{\tau}
\left(\tau+\int_{0}^{t}d\xi\;e^{\xi/\tau}\;\mid
M(\xi/2)\mid^2\right)\;.
\end{equation}
For the free induction decay being given by
$M(t)\approx\sum_{\nu=1}^N f_\nu\exp(-\Gamma_\nu t)$ one obtains
\begin{equation}
\label{mtc} {M_{SE}}(t)= \sum_{\mu,\nu=1}^N\frac{f_\mu f_\nu
}{1-\tau R_{\mu\nu}} \left(e^{-R_{\mu\nu}t} - \tau
R_{\mu\nu}e^{-t/\tau} \right) \;, \label{solution2}
\end{equation}
where $R_{\mu\nu}=(\Gamma_\mu+\Gamma_\nu)/2$.

The determination of the relaxation time $T_2$ of ${M_{SE}}(t)$
depends on the experimental setup. When $T_2$ is determined from a
single echo time $t$ or a multi-echo sequence \cite{gill} with an
inter-echo time $t$, then $T_2$ is usually obtained from
\begin{equation}
\label{T21} T_2=-\ln({M_{SE}} (t))/t \;.
\end{equation}
However, since ${M_{SE}}(t)$ is $not$ single exponential already
in the simplest case, $N=1$, the above definition renders $T_2$
effectively dependent on the echo time $t$.

When $T_2$ is determined from several experiments with varying
echo times the best single exponential approximation,
${M_{SE}}(t)\approx e^{-t/T_2}$, is required. Within the mean
relaxation time approximation, $T_2=\int_0^{\infty} dt
{M_{SE}}(t)/{M_{SE}}(0)$, Eq.~(\ref{solution2}) gives
\begin{equation}
\label{T22} T_2=T_2^*+\tau
\end{equation}

Equations (\ref{T2star}) and (\ref{T22}) allow the analysis of
various limiting cases. The {\it motional narrowing} regime is
given for $\tau\langle\omega^2\rangle^{1/2}\ll 1$. Then, the
relaxation rate of the free induction process is
$1/T_2^*=\tau\langle\omega^2(x)\rangle$ with the spatial variance
of the spin precession frequency $\langle\omega^2(x)\rangle=
\varsigma\delta\omega^2/2$. This relation is the well known
general result for the transverse relaxtion rate obtained for the
motional narrowing limit.

For very long correlation times one obtains
$1/T_2^*=2\varsigma\delta\omega$ and $1/T_2=0$, i.e relaxation is
due solely to coherent spin dephasing. For very short correlation
times one obtains $T_2^*\approx T_2$, i.e. in this limit the free
induction decay is mainly due to incoherent spin dephasing.

Another approximation is possible for small relative
intracapillary blood volumes ($\varsigma\ll 1$) and in the regime
$\tau\delta\omega\le 1$, which is realistic in cardiac tissue.
Equation (\ref{T2star}) then provides $1/T_2^*=\varsigma
\tau^{-1}(\sqrt{1+(\tau\delta\omega)^2}-1)$, i.e. the rate of the
free induction decay is proportional to the intracapillary volume.

Kennan and Gore  \cite{kennan} determined relaxation rates from
{\it Monte Carlo simulations} of transverse polarized spins as a
function of the diffusion coefficient. In Figs. 1 and 2 their
results are compared to our theoretical predictions from
Eqs.~(\ref{T2star}), (\ref{mtc}),
 and (\ref{T21}),
using their tissue parameters and $N=1$, which proved to be a
sufficient approximation. They demonstrate a close similarity of
analytically obtained  $T_2^*$ $vs$ $D$ and $T_2$ $vs$ $D$ curves,
and those of simulations. The correct asymptotic behaviour for
very long and short correlation times, as derived above, is
evident in both figures. Also the location of the maximum of the
$T_2$ $vs$ $D$ curves is congruent. Figure 1 demonstrates
furthermore that the dependence of $T_2$ on the echo time is
described correctly. Figure 2 exhibits the correct dependence on
the capillary radius.

We recently performed $T_2^*$ mapping in hearts of volunteers
\cite{wacker2} before and after application of the coronary
dilator dipyridamol, which increases coronary flow by a factor of
about 5 without affecting oxygen consumption. This decreases the
deoxyhemoglobin content and one observes an increase of
$T_2^*\approx 36 {\rm ms}$ by 17\%. The theory presented above
predicts an increase of 13 to 18\%, which is in good agreement
with the experimental data.

To summarize, we developed an analytical theory for the
description of spin dephasing in the inhomogeneous magnetic field
of the capillary network of myocardium. The underlying strong
collision approach is justified in the cardiac muscle and, most
probably,  albeit with modifications according to different
vascular architectures, for many other tissues as well. Note that
this approach is not only valid when spin dephasing occurs on a
much longer time scale than the local field fluctuations, which is
the basis for the strong collison approximation, but also in the
static dephasing regime.

In our applications we considered only the single-exponential
(mean relaxation time) approximation of the full relaxation curve
$M(t)$, which proved sufficient for the parameter regime in
question. However, in particular for short spin echo times, it can
be important to describe the short time behavior of $M(t)$, e. g.
$\frac{d}{dt} M(t)|_{t=0}=0$, in more detail. This is readily
possible using the multi-exponential approximation for $M(t)$, see
also \cite{bauer99c}.

This work was supported by the Deutsche Gesellschaft f\"ur
Kardiologie, Forschungsfonds des Klinikum Mannheim/Heidelberg
Projekt 42, Grant Sonderforschungsbereich 355 ``Pathophysiologie
der Herzinsuffizienz'', and Graduiertenkolleg ``NMR'' HA 1232/8-1.

\begin{figure}
\vskip -2.5truecm \epsfig{file=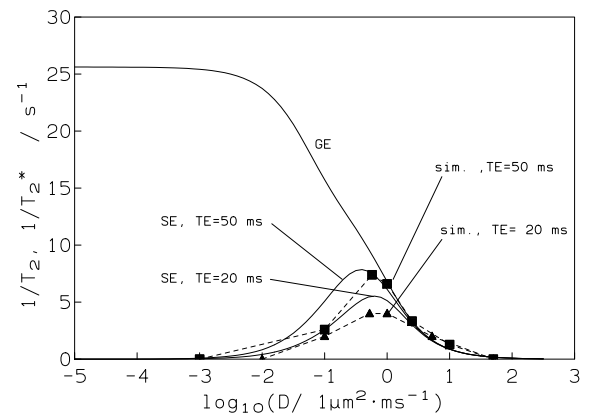,width=8.5truecm} \vskip
-2.5truecm \caption{ Relaxation rates of the free induction decay
of a gradient echo ($1/T_2^*$, GE) and spin echo experiment
($1/T_2$, SE) as function of the diffusion coefficient. For
comparison with simulation data (sim.), the tissue parameters were
obtained from Ref. \protect\cite{kennan} as: relative
intracapillary blood volume $\varsigma=5\%$, capillary radius
$R_c=2.5\mu$m, intracapillary magnetization of 1.6 mgauss
($\delta\omega=269 {\rm rad\; s}^{-1}$).  The theoretical results
(solid lines) were obtained from Eq.~(\protect{\ref{T2star}}) for
the free induction decay,and from Eqs.(\protect\ref{mtc}) and
(\protect\ref{T21}) for the spin echo. Two echo times (TE) were
analyzed. \label{fig:fig1} }
\end{figure}

\begin{figure}
\vskip -2.5truecm \epsfig{file=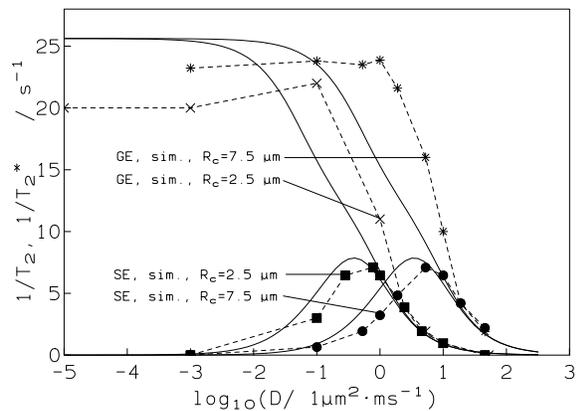,width=8.5truecm} \vskip
-2.5truecm \caption{ Relaxation rates of free induction decay
($1/T_2^*$, GE) and spin echo experiments ($1/T_2$, echo time$=50$
ms, SE). Simulation (sim.) and theoretical (solid lines)results
are compared for a capillary radius of $2.5\mu$m and $7.5\mu$m.
Other parameters are as in Fig. 1. \label{fig:fig2} }
\end{figure}

\end{multicols}

\end{document}